\documentstyle[epsf,algo,12pt]{article}
\newcommand{\eqnref}[1]{(\ref{#1})}

\newcommand{\one}{\mbox{{\bf 1}}}
\newcommand{\beq}{\begin{equation}}
\newcommand{\eeq}{\end{equation}}        
\newcommand{\bqa}{\begin{eqnarray}}        
\newcommand{\eqa}{\end{eqnarray}}        
\newcommand{\be}{\begin{enumerate}}
\newcommand{\ee}{\end{enumerate}}        
\newcommand{\bi}{\begin{itemize}}
\newcommand{\ei}{\end{itemize}}        

\newcommand{\eg}{{\frenchspacing\em e.\hspace{0.4mm}g.{}}}
\newcommand{\ie}{{\frenchspacing\em i.\hspace{0.4mm}e.{}}}

\title{{\small \hfill WUB 96-1}\\[-.2cm]
       {\small \hfill HLRZ 4/96}\\[1cm]
A Parallel SSOR Preconditioner for Lattice QCD}
\author{S. Fischer$^{a}$, A. Frommer$^{a}$, U. Gl\"assner$^{b}$,\\ 
        Th. Lippert$^{c}$, G. Ritzenh\"ofer$^{c}$,
        and K. Schilling$^{b,c}$\\[0.4cm]
        \footnotesize
        $^{a}$Department of Mathematics, University of Wuppertal,
        42097 Wuppertal, Germany\\[-.1cm]
        \footnotesize
        $^{b}$Department of Physics, University of Wuppertal,
        42097 Wuppertal, Germany\\[-.1cm]
        \footnotesize
        $^{c}$HLRZ, c/o KFA-J\"ulich, D-52425 J\"ulich,
        Germany and DESY, Hamburg, Germany}
\date{ }
\begin{document}
\maketitle

\begin{abstract}{We present a parallelizable SSOR preconditioning scheme for
  Krylov subspace iterative solvers which proves to be efficient in
  lattice QCD applications involving Wilson fermions. Our
  preconditioner is based on a locally lexicographic ordering of the
  lattice points.  In actual hybrid Monte
  Carlo applications with the bi-conjugate gradient stabilized method
  BiCGstab,
  we achieve a gain factor of about 2 in the number of
  iterations compared to conventional odd-even preconditioning.
  Whether this translates into similar reductions in run time will
  depend on the parallel computer in use.  We discuss implementation
  issues using the `Eisenstat-trick' 
  and machine specific advantages of the method for the
  APE100/Quadrics parallel computer.  In a full QCD simulation on a
  512-processor Quadrics QH4 we find a gain in cpu-time of a factor of
  1.7 over odd-even preconditioning for a $24^3\times 40$ lattice.}
\end{abstract}
  
\parindent 0pt
\parskip 12 pt

\section{Introduction} The computation of the effects of fermionic
forces (from virtual quark-antiquark creation and annihilation
processes) onto the vacuum structure of quantum chromodynamics (QCD)
presents a severe bottleneck to the numerical evaluation of this
fundamental theory of strong interactions.
 
The stochastic sampling of the vacuum state by the standard Hybrid
Monte Carlo algorithm (HMC) \eg\ implies the continual computation of
fermionic Greens functions on a stochastic source $\phi$. In terms of
the discretized Dirac operator $M$, the fermionic Greens function (or
`quark propagator') is nothing but the solution of the linear equation
\beq Mx=\phi, \label{Wilson_eq} \eeq where $M$ is a huge sparse
matrix, of rank $r = 3\times 4 \times V$ and $V$ is the volume of the
underlying 4-dimensional space-time lattice. There are different
discretizations of the Dirac operator in use. For the study of weak
interaction processes, the Wilson fermion scheme \cite{WILSON} appears
to be the most attractive.

Today, parallel computers have matured and provide sufficient compute
power to drive simulations of full QCD with Wilson fermions to the
level of 100 and more Teraflops-hours.  This performance allows for
sufficient sampling of QCD vacuum configurations on reasonably sized
lattices.  As a consequence there is topical interest to {(\em i)}
pioneer new algorithms for generating QCD configurations on one hand
\cite{Lu93,DEFORCRAND,JEGERLEHNER} and {\em (ii)} achieve progress
in improving HMC (as the method of choice) on the other
hand\footnote{It appears promising to find that this complementary
  strategy of tackling the fermion problem in numerical field theory
  joins both physicists and applied mathematicians.}.

Meanwhile, the stabilized bi-conjugate gradient algorithm (BiCGstab)
\cite{vV92} has been established as an efficient inverter in lattice
QCD applications \cite{FHNLS94,BF94,FHNLS94_2,dF95} since it requires
less iterations and less computing time as compared to the minimal
residual algorithm or CG on the normal equations: improvements of
about 50 \% (with respect to the conjugate gradient) can be achieved
in the regime of small quark masses. Numerical studies in
Refs.~\cite{FHNLS94,BF94,Fr94} indicate that further progress is
now to be expected through preconditioning.

In the past, various approaches for preconditioning the Dirac matrix
$M$ have been taken:
\begin{enumerate}
\item The first attempt started from the perception that in the weak
  coupling limit the quark propagator is diagonal in momentum space
  and hence Fourier transformation would act as a useful
  preconditioner.  Such `Fourier acceleration' \cite{DAVIES},
  however, needs smoothness of fields and hence gauge fixing. This
  handicaps the computation with a substantial overhead which at the
  end of the day eats up most of the previous gain.
\item The state-of-the-art\footnote{Non-Krylov-subspace based methods
    like multigrid as of today did not achieve a level of maturity to
    be competitive with inverters like CG in lattice gauge theory.
    This is due to the disordered nature of the gluonic field
    configurations that act as fluctuating coefficients within the
    discretized differential operator stencils.} preconditioning
  approach in lattice QCD\footnote{The transformation of $M$ from
    natural into odd-even order might be looked upon as a particular
    SSOR preconditioning, see Section~\ref{ORDERINGS}.} rests upon the
  odd-even decomposition of the matrix $M$ \cite{ROSSI} that yields an
  efficiency gain by a factor of 2 - 3 when inverting $M$ in the
  context of actual fermionic force and quenched propagator
  computations \cite{GUPTA,FHNLS94}. It is crucial that this second
  preconditioning approach lends itself easily to parallelization.
\item A third and promising method has been evaluated by Oyanagi in
  the 80s \cite{Oy85}.  He used the standard incomplete LU (ILU)
  factorization of the matrix $M$ based on the natural, {\em globally
    lexicographic}\/ ordering of the lattice points. For Wilson
  fermions with Wilson parameter $r=1$, the ILU preconditioner is
  identical to the SSOR preconditioner with respect to that ordering.
  
  Ref.~\cite{Oy85} found ILU preconditioning (for the conjugate
  residual method) to outperform the odd-even decomposition on vector
  machines, achieving a large gain factor (in terms of iteration
  numbers) over conventional unpreconditioned methods. As it stands,
  Oyanagi's method works satisfactorily on vector machines. However,
  on local memory or grid-oriented parallel computers, this
  preconditioner can hardly be implemented efficiently, since
  parallelism can only be achieved by working concurrently on lattice
  points lying on the same {\em diagonal hyper-plane}. Hockney
 \cite{HOCKNEY,HOCKNEY2}
  reports that the run times on several parallel and vector computers
  actually degrade as compared to the odd-even preconditioning.

  Using the so-called `Eisenstat-trick' (see Section~\ref{SSOR}),
  the ILU preconditioning can be implemented in a more efficient manner
  than in \cite{Oy85,HOCKNEY}\footnote{Incorporating the Eisenstat trick
  into the Oyanagi ILU 
  preconditioner could thus lead to a revision of some of the conclusions
  from \cite{HOCKNEY}, so that a more detailed numerical study is advisable.}. 
\end{enumerate}

In this paper, which is based on the results in \cite{Fi96},
we intend to open the stage for use of general parallel
SSOR preconditioning techniques in lattice QCD based on appropriate
orderings of the lattice points. Our approach may be regarded as a
generalization of the odd-even (or red-black) ordering to a multiple
color layout \cite{FROMMER:NUM} or, alternatively, as a localization
of the globally lexicographic ordering. The parallel implementation
will be carried out in combining two steps:
\begin{enumerate}
\item Partition the whole lattice into equally shaped sublattices
and introduce a lexicographic ordering on each sublattice.
\item During the SSOR preconditioning step, in parallel sweep 
through all sublattices in lexicographic order. 
\end{enumerate}

In Section \ref{SSOR} we discuss the SSOR preconditioner in
detail and we give efficient implementations for the
SSOR preconditioned BiCGstab and MR methods involving the
Eisenstat-trick. Section \ref{ORDERINGS} explains how
different orderings of the lattice points lead to variants of the SSOR
preconditioning procedure.  In particular, this section introduces
the locally lexicographic ordering through which we obtain
the improvements over odd-even preconditioning.

In Section \ref{IMPLEMENTATION}, the parallel
implementation in conjunction with sub-blocking is discussed.

Section \ref{RESULTS} presents the behavior of our new
preconditioning method for Hybrid Monte Carlo applications and its
performance on APE100/Quadrics parallel computers.  On lattices of
size $8^3\times 16$ we test the efficiency of the method numerically
where we find improvement factors of about 2 in terms of iteration
numbers compared to the standard odd-even method.  We discuss the
effects of granularity and further report about first experiences with
the preconditioner in a large scale simulation using a 512-node
APE100/Quadrics on a $24^3\times 40$ lattice, in the chirally sensible
region.

\section{SSOR Preconditioning\label{SSOR}}

When preconditioning \eqnref{Wilson_eq}, we take two non-singular
matrices $V_1$ and $V_2$ which act as a left and a right
preconditioner, respectively, i.e.\ we consider the new system
\begin{equation}
V_1^{-1}MV_2^{-1} \tilde{x} = \tilde{\phi}, \enspace \mbox{where } 
\tilde{\phi} = V_1^{-1}\phi,
\; \tilde{x} = V_2x. \label{Wilson_prec_eq}
\end{equation}
(In the following, preconditioned quantities will be denoted by
tildes.)  We could now apply BiCGstab, as given in Algorithm~\ref{BiCGstab}
for the original system, (or any other Krylov subspace method) directly
to \eqnref{Wilson_prec_eq}, replacing each occurrence of $M$ and $\phi$
by $V_1^{-1}MV_2^{-1}$ and $\tilde{\phi}$, resp.
\begin{algor}[htb]
\begin{center}
\begin{minipage}{\textwidth}
\begin{tabbing}
\hspace*{2ex} \= \hspace{2ex} \= \kill
 \{ initialization \}  \\  
  choose ${x}_{0}$, set ${r}_{0}={\phi}-{M}{x}_{0}, \hat{r}_{0}={r}_{0}$ \\ 
 set  $ {\rho}_{0}={\rho}_{1}={\alpha}_{0}={\omega_{0}}=1$ \\ 
 set  $ {v}_{0}={p}_{0}=0$ \\
 \{ iteration \} \\
 for  $i =1, 2, \ldots$ \\   
 \> ${\rho}_{i}= \hat{r}_{0}^{\dagger}{r}_{i-1} $ \\  
 \> ${\gamma}_{i} = ({\rho}_{i}/{\rho}_{i-1})({\alpha}_{i-1}/{\omega}_{i-1})$ \\
 \> ${p}_{i} = {r}_{i-1}+{\gamma}_{i}({p}_{i-1}-{\omega}_{i-1}{v}_{i-1}) $\\
 \> ${v}_{i}={M}{p}_{i}$ \\
 \> ${\alpha}_{i}={\rho}_{i}/  \hat{r}_{0}^{\dagger}{v}_{i}$ \\  
 \> ${s}_{i} = {r}_{i-1}-{\alpha}_{i} {v}_{i}$ \\
 \> ${t}_{i}= {M}{s}_{i}$ \\  
 \> ${\omega}_{i}={t}_{i}^{\dagger} {s}_{i} / {t}_{i}^{\dagger} {t}_{i}$ \\ 
 \> ${x}_{i} ={x}_{i-1}+{\omega}_{i}{s}_{i}+{\alpha}_{i}{p}_{i}$ \\ 
 \> ${r}_{i}={s}_{i}-{\omega}_{i}{t}_{i}$   
\end{tabbing}
\end{minipage}
\end{center}
\caption[dummy]{BiCGstab method\label{BiCGstab}}
\end{algor}
However, this would yield the preconditioned iterates
$\tilde{x}^k$ together with preconditioned residuals. Therefore, one usually
reformulates the algorithm incorporating an implicit back-transformation to the
unpreconditioned quantities. The resulting algorithm is similar to
Algorithm~\ref{BiCGstab}, requiring two
additional systems with matrix $V = V_1V_2$ and two systems with matrix $V_1$ 
to be solved in each iterative step (see \cite{vV92}).  

The purpose of preconditioning is to reduce the number of iterations
and the computing time necessary to achieve a given accuracy. This means that
$V$ has to be a sufficiently good approximation to the inverse of $M$
(thus decreasing the number of iterations) while solving systems with
$V,V_1$ should be sufficiently cheap (since these solves
represent the overhead of the preconditioned upon the basic
method).

In the present work we are only interested in the SSOR (or, more precisely,
symmetric Gau\ss{}-Seidel) preconditioner. Consider the decomposition
of $M$ into its diagonal, strictly lower and strictly upper triangular parts,
i.e.\
\[
M = I - L - U
\]
with $L,U$ strictly lower and upper triangular matrices,
respectively. Then the SSOR preconditioner is given by 
\begin{equation}
\label{SSOR_prec_eq}
V_1 = I - L, \enspace V_2 = I-U.
\end{equation}

\begin{algor}[t]
\begin{center}
\begin{minipage}{\textwidth}
\begin{tabbing}
\hspace*{2ex} \= \hspace{35ex} \= \kill
 \{ initialization \}  \\
  choose ${x}_{0}$, set $r_0 = \phi - Mx^0$ \\
  solve $(I-L)\tilde{r}_0 = r_0$ to get $\tilde{r}_0$ \> \>
                   \{ forward solve \} \\
  $\tilde{\hat{r}}_{0}=\tilde{r}_{0}$ \\
 set  $ {\rho}_{0}={\rho}_{1}={\alpha}_{0}={\omega_{0}}=1$ \\
 set  $ \tilde{v}_{0}=\tilde{p}_{0}=0$ \\
\{ iteration \} \\
 for  $i =1, 2, \ldots$ \\
 \> ${\rho}_{i}= \tilde{\hat{r}}_{0}^{\dagger}\tilde{r}_{i-1} $ \\
 \> ${\gamma}_{i} = ({\rho}_{i}/{\rho}_{i-1})({\alpha}_{i-1}/{\omega}_{i-1})$
\\ \> $\tilde{p}_{i} = \tilde{r}_{i-1}+{\gamma}_{i}(\tilde{p}_{i-1}-
    {\omega}_{i-1}\tilde{v}_{i-1}) $\\
 \> solve $(I-U){z}_i = \tilde{p}_i $ to get $z_i$ \> \{ backward solve \} \\
 \> solve $(I-L)\tilde{w}_i = \tilde{p}_i - {z}_i$ to get $\tilde{w}_i $ \>
     \{ forward solve \}\\
 \> $\tilde{v}_i = z_i + \tilde{w}_i$ \\
 \> ${\alpha}_{i}={\rho}_{i}/  \tilde{\hat{r}}_{0}^{\dagger}\tilde{v}_{i}$ \\
 \> $\tilde{s}_{i} = \tilde{r}_{i-1}-{\alpha}_{i} \tilde{v}_{i}$ \\
 \> solve $ (I-U){y}_i = \tilde{s}_i $ to get $y_i$\> \{ backward solve \} \\
 \> solve $(I-L)\tilde{u}_i = {s}_i - {y}_i $ to get $\tilde{u}_i$
        \> \{ forward solve \} \\
 \> $\tilde{t}_i = y_i + \tilde{u}_i$   \\
 \> ${\omega}_{i}=\tilde{t}_{i}^{\dagger} \tilde{s}_{i} /
    \tilde{t}_{i}^{\dagger} \tilde{t}_{i}$ \\
 \> ${x}_{i} ={x}_{i-1}+{\omega}_{i}{y}_{i}+{\alpha}_{i}{z}_{i}$ \\
 \> $\tilde{r}_{i}=\tilde{s}_{i}-{\omega}_{i}\tilde{t}_{i}$
\end{tabbing}
\end{minipage}
\end{center}
\caption[dummy]{SSOR preconditioned BiCGstab\label{SSOR-BiCGstab}}
\end{algor}

For the SSOR preconditioner we have 
$V_1 + V_2 - M  = I$. This important relation
can be exploited through the `Eisenstat-trick' \cite{Eis81}, 
because we
now have $V^{-1}MV_2^{-1} = V_2^{-1} + V_1^{-1}(I-V_2^{-1})$, so that the
matrix vector product $w = V_1^{-1}MV_2^{-1}r$ can economically be computed via
\[
v = V_2^{-1}r, \enspace u = V_1^{-1}(r-v), \enspace w = v + u.
\]
In this manner we get the
standard algorithmic formulation of the SSOR preconditioned BiCGstab method,
stated as Algorithm~\ref{SSOR-BiCGstab}.

The Eisenstat trick is not restricted to the BiCGstab method but can be applied
in any Krylov subspace method. As another
example, we state the SSOR preconditioned
minimal residual (MR) method as Algorithm~\ref{MR}.

\begin{algor}[thb]
\begin{center}
\begin{minipage}{\textwidth}
\begin{tabbing}
\hspace*{2ex} \= \hspace{35ex} \= \kill
 \{ initialization \}  \\
  choose ${x}_{0}$, set $r_0 = \phi - Mx^0$ \\
  solve $(I-L)\tilde{r}_0 = r_0$ to get $\tilde{r}_0$ \> \>
                              \{ forward solve \} \\
\{ iteration \} \\
 for  $i =0, 1, \ldots$ \\
 \> solve $(I-U)w_i = \tilde{r}_i$ to get $w_i$ \> \{ backward solve \} \\
 \> solve $(I-L)v_i = \tilde{r}_i - w_i$ to get $v_i$ \> \{ forward solve \} \\
 \> $\tilde{p}_i = w_i + v_i$ \\
 \> $\alpha_i =
         \tilde{r}_i^{\dagger}\tilde{p}_i /\tilde{p}_i^{\dagger}\tilde{p}_i$ \\
 \> $ x_{i+1} = x_i + \alpha_i w_i$ \\
 \> $\tilde{r}_{i+1} = \tilde{r}_i - \alpha_i\tilde{p}_i$
\end{tabbing}
\end{minipage}
\end{center}
\caption[dummy]{SSOR preconditioned minimal residual\label{MR}}
\end{algor}

Note that these algorithms use the preconditioned residuals
$\tilde{r}_i$ which are related to the unpreconditioned residuals $r_i
= \phi - M x_i$ via
\[
{r}_i = (I-L)\tilde{r}_i.
\]
To save computational costs, a stopping criterion for
Algorithm~\ref{SSOR-BiCGstab} will usually be based on $\tilde{r}_i$.
Upon successful stopping, one can then compute $r_i$ and test for
convergence using $r_i$. If the solution is not yet accurate enough,
one continues the iteration with a stronger stopping criterion. In our
numerical experiments to be reported in Section~\ref{RESULTS} it
turned out that such additional steps were never necessary.

Also note that multiplications with $M$ are completely
absent in this formulation, the only matrix operations being
solves with each of the matrices $I-L$ and $I-U$.
Since these matrices are triangular, the solves can be done directly
via forward or backward substitution, respectively. For example, denoting
the components of a vector $y$ by $(y)_i$ and the  entries of $L$ by $(L)_{ij}$
and similarly for $U$, the
forward solve $(I- L)y = p$ and backward solve $(I-U)z = p$ become
simply
 
\begin{center}
\begin{minipage}{5cm}
\begin{tabbing}
\hspace*{2ex} \= \hspace{2ex} \= \kill
for $i=1,\ldots,n$ \\
\> $\displaystyle (y)_i = (p)_i + \sum_{j=1}^{i-1} (L)_{ij}(y)_j$
\end{tabbing}
\begin{center} forward solve \end{center}
\end{minipage}
\hspace*{2cm}
\begin{minipage}{5cm}
\begin{tabbing}
\hspace*{2ex} \= \hspace{2ex} \= \kill
for $i=n,\ldots,1$ \\
\> $\displaystyle (z)_i = (p)_i + \sum_{j=i+1}^{n}(U)_{ij}(z)_j$
\end{tabbing}
\begin{center} backward solve \end{center}
\end{minipage}
\end{center}

Due to the sparsity pattern of $M$, most of the entries in $L$ and $U$ are
zero so that only a few $j$ actually contribute to the sums over $j$.
We will come back to this point in more detail in the next section.

It is important to note that for $i$ fixed, the number of non-zero
entries of $L$ and $U$ involved when updating $y_i$ in the forward together
with $z_i$ in the backward solve is just the number of non-zero entries of
the $i$-th row of the matrix $M$. Therefore, in terms of computational cost,
a forward followed by a backward solve is quite precisely as
expensive as a
multiplication with $M$ (which is required in the unpreconditioned
method). Hence, due to the Eisenstat-trick, 
in terms of matrix vector operations,
each step of the SSOR preconditioned BiCGstab (or MR)
method requires the same work as in the unpreconditioned method.

We finally note that the `classical' symmetric Gau\ss{}-Seidel
iteration
\[
(I-L)x^{k+1/2} = Ux^{k}+ \phi, \enspace (I-U)x^{k+1} = L x^{k+1/2} + \phi, 
\enspace k = 0,1,\ldots
\]
represents a sequence of alternating forward and backward solves.
This explains the terminology SSOR preconditioner where SSOR stands for
symmetric successive over-relaxation, the over-relaxed variant of the
symmetric Gau\ss{}-Seidel method. See \cite{Or90}, e.g.

\section{Orderings\label{ORDERINGS}}

When writing down equation \eqnref{Wilson_eq} with Wilson fermion
matrix $M$\footnote{The 4 matrices $\gamma_{\mu}$ are $4\times 4$
  Dirac matrices, and the $U_{\mu}$ are $3\times 3$ SU(3) matrices
  that represent the gluonic degrees of freedom on the lattice.}
\bqa
M_{x,y}&=& \delta_{x,y} -\kappa\sum_{\mu=1}^4 (1-\gamma_{\mu})U_{\mu}(x)\,
           \delta_{x,y-{\mu}} + (1+\gamma_{\mu})U^{\dagger}_{\mu}(x-\mu)\,
           \delta_{x,y+{\mu}}\nonumber\\
       &=& \delta_{x,y} -\kappa\sum_{\mu=1}^4
       m^-_{x,y}\,\delta_{x,y-{\mu}}
     + m^+_{x,y}\, \delta_{x,y+{\mu}}.
\eqa 
we have the freedom to choose any ordering scheme for the lattice
points $x$. Different orderings yield different matrices $M$, however,
which are permutationally similar to each other, i.e.\ one matrix can
be retrieved from the other via the transformation $M \rightarrow
P^{\dagger}MP$ with a permutation matrix $P$.  In general,
$P^{\dagger}LP$ and $P^{\dagger}UP$ will {\em not} represent the
strictly lower or strictly upper triangular part of $P^{\dagger}MP$.
Consequently, the SSOR preconditioned matrix for $P^{\dagger}MP$ will
{\em not} be permutationally similar to that of $M$, so that the
quality of the SSOR preconditioner will usually depend on the ordering
scheme chosen. One purpose of this paper is to explain that the
odd-even preconditioning and the Oyanagi preconditioning can both be
regarded as the SSOR preconditioning belonging to particular orderings
of the grid points\footnote{It is shown in \cite{Fi96} that the SSOR
preconditioner for the Wilson fermion matrix (with Wilson parameter $r=1$)
is identical to the ILU preconditioner for {\em any} ordering of the 
lattice points}. We will then introduce a new `locally
lexicographic' ordering which is adapted for parallel computation and
for which the SSOR preconditioned system can be solved more rapidly
than with odd-even preconditioning with respect to both, the number of
iterations as well as actual run time on particular parallel
computers.

Consider an arbitrary numbering (ordering) of the lattice points. For
a given grid point $x$, the corresponding row in the matrix $L$ or $U$
contains exactly the coupling coefficients of those nearest neighbors
of $x$ which have been numbered before or after $x$, resp.  Therefore,
a generic formulation of the forward solve for this ordering is given
by Algorithm~\ref{generic-forward}.  The backward solves are done
similarly, now running through the grid points in {\em reverse} order
and taking those grid points $x \pm \mu$ which were 
numbered {\em after} (instead of {\em before}) $x$. 
Due to this analogy, we will restrict our
discussion to the forward solves in the sequel.

\begin{algor}[htb]
\begin{center}
\begin{minipage}{\textwidth}
\begin{tabbing}
\hspace*{2ex} \= \hspace{2ex} \=  \hspace{2ex} \= \hspace{2ex} \kill
for all grid points $x$ in the given order \\
\> \{ update $y_x$ \} \\
\> $y_x = p_x $ \\
\> for $\mu = 1,\ldots,4$ \\
\>  \> if $x-\mu$ was numbered before $x$ then \\
\>  \>  \> $\displaystyle y_x = y_x + \kappa \cdot  m^+_{x,x-\mu}y_{x - \mu}$ \\
\> for $\mu = 1,\ldots,4$ \\
\>  \> if $x+\mu$ was numbered before $x$ then \\
\>  \>  \> $\displaystyle y_x = y_x + \kappa \cdot  m^-_{x,x+\mu}y_{x + \mu}$ 
\end{tabbing}
\end{minipage}
\end{center}
\caption{Generic forward solve\label{generic-forward}}
\end{algor}

\paragraph{Odd-even ordering.} 
As a first specific example, let us consider the familiar odd-even
ordering where all odd lattice points are numbered before the even
ones.  From our generic formulation in
Algorithm~\ref{generic-forward} we then get
Algorithm~\ref{odd-even-forward} for the forward solve.

\begin{algor}[htb]
\begin{center}
\begin{minipage}{\textwidth}
\begin{tabbing}
\hspace*{2ex} \= \hspace{2ex} \=  \hspace{2ex} \= \hspace{2ex} \kill
for all odd grid points $x$ \\
 \> $y_x = p_x$ \\
for all even grid points $x$ \\
\> \{ update $y_x$ \} \\
\>  $\displaystyle y_x = p_x + \kappa \left(
         \sum_{\mu= 1}^{4} m^+_{x,x-\mu}\, y_{x - \mu}
         + \sum_{\mu = 1}^{4} m^-_{x,x+\mu}\, y_{x+\mu}
              \right)$
\end{tabbing}
\end{minipage}
\end{center}
\caption{Odd-even forward solve\label{odd-even-forward}}
\end{algor}

In traditional QCD computations, the odd-even preconditioning is {\em
  not} implemented by using the above formulation of the forward (and
backward) solve in Algorithm~\ref{SSOR-BiCGstab}.  This is due to the
fact that---very exceptionally---for this particular ordering the
inverses of $I-L$ and $I-U$ can be determined directly: With the
odd-even ordering, the matrix $M$ has the form
\begin{equation}\label{Wilson_odd_even_eq}
M = \left(
    \begin{array}{cc}
      \one & - \kappa D_{oe} \\
     - \kappa D_{eo}              &  \one
     \end{array}
    \right)  
\end{equation}
so that
\[
I-L = \left(
    \begin{array}{cc}
      \one & 0 \\
     - \kappa D_{eo}  &  \one
     \end{array}
    \right)
\mbox{ with }
(I-L)^{-1} = \left(
    \begin{array}{cc}
      \one & 0 \\
      \kappa D_{eo}  &  \one
     \end{array}
    \right)
\]
and
\[
I-U = \left(
    \begin{array}{cc}
      \one & - \kappa D_{oe} \\
     0              &  \one
     \end{array}
    \right)
\mbox{ with }
(I-U)^{-1} = \left(
    \begin{array}{cc}
      \one &  \kappa D_{oe} \\
     0              &  \one
     \end{array}
    \right) .
\]
Hence
\[
(I-L)^{-1}M(I-U)^{-1} =
\left(
    \begin{array}{cc}
     \one & 0 \\
     0              & \one -\kappa^2 D_{eo}D_{oe}
     \end{array}
    \right),
\]
where $\one - \kappa^2 D_{eo}D_{oe}$ is called the matrix of the
odd-even reduced system.  Very exceptionally again, in this
particular case, the preconditioned matrix $(I-L)^{-1}M(I-U)^{-1}$
thus has such a simple structure that it is possible to apply
Algorithm~1 directly, using the preconditioned matrix explicitly.
Since the matrix $(I-L)^{-1}M(I-U)^{-1}$ is $2\times 2$ block
diagonal with the first diagonal block being the identity, the
computations for the second diagonal block are completely decoupled
from those of the first block and they are identical to those which
are performed when applying the algorithm for the odd-even reduced
system directly. This is precisely what is done in traditional
odd-even preconditioning. So traditional odd-even preconditioning is
equivalent to SSOR preconditioning for the odd-even ordered system.
 
So far, odd-even preconditioning is generally considered as the only
successful preconditioner in a parallel computing environment. For
typical, realistic configurations, it gains a factor 2-3 in the
numbers of iterations and also in computing time as compared to
solving the unpreconditioned system.

\paragraph{Globally lexicographic ordering.} 
Assume now that $M$ is given with respect to the natural
(lexicographic) ordering of the lattice points. This means that grid
point $x = (i_1,i_2, i_3,i_4)$ is numbered before $x' =
(i_1',i_2',i_3',i_4')$ if and only if ($i_4 < i_4'$) or ($i_4 = i_4'$
and $i_3 < i_3'$) or ($i_4 = i_4', i_3 = i_3'$ and $i_2 < i_2'$) or
($i_4 = i_4', i_3 = i_3', i_2 = i_2'$ and $i_1 < i_1'$).  The
corresponding forward solve is given as Algorithm~\ref{lex-forward}.

\begin{algor}[htb]
\begin{center}
\begin{minipage}{\textwidth}
\begin{tabbing}
\hspace*{2ex} \= \hspace{2ex} \=  \hspace{2ex} \= \hspace{2ex} \= \hspace{2ex}
   \= \hspace{2ex} \= \kill
for $i_4 = 1,\ldots,n_4$  \\
\> for $i_3 = 1,\ldots,n_3$  \\
\>  \> for $i_2 = 1,\ldots,n_2$  \\
\>  \>  \> for $i_1 = 1,\ldots,n_1$  \\
\>  \> \> \> $x := (i_1,i_2,i_3,i_4)$ \\
\>  \> \>  \> \{ update $y_x$ \} \\
\>  \> \>  \> $y_x = p_x $ \\
\>  \> \>  \> for $\mu = 1,\ldots,4$ \\
\>  \> \> \>  \> if $i_{\mu} > 1$ then $y_x = y_x +
                        \kappa\, m^+_{x,x-\mu}y_{x-\mu}$ 
\end{tabbing}
\end{minipage}
\end{center}
\caption{Lexicographic forward solve \label{lex-forward}}
\end{algor}

The SSOR preconditioning for the lexicographic ordering, applied to the
MR and other conjugate residual methods, was considered by Oyanagi \cite{Oy85}. 
He showed that it yields a
further improvement over odd-even preconditioning as far as the number
of iterations is concerned.  However, its parallel implementation is
more difficult and less efficient, since only grid points lying on the
same {\em diagonal hyper-plane} can be worked upon in parallel in the
forward and backward solves.
Hockney \cite{HOCKNEY2} reports that on the 592 processor
ACPMAPS at Fermi-Lab, the run time (without the Eisenstat trick)
actually degrades as compared to odd-even preconditioning. 
See also \cite{HOCKNEY}.

\paragraph{Locally lexicographic ordering.} 
Unlike the lexicographical and the odd-even ordering, the ordering we
propose now is adapted to the parallel computer used to solve equation
\eqnref{Wilson_eq}.  We assume that the processors of the parallel
computer are connected as a $p_1 \times p_2 \times p_3 \times p_4$
4-dimensional grid. Note that this includes lower dimensional grids by
setting some of the $p_i$ to 1. The total number of processors is $p =
p_1p_2p_3p_4$. The space-time lattice can be matched to the processor
grid in an obvious natural manner, producing a local lattice of size
$n^{loc}_1 \times n^{loc}_2 \times n^{loc}_3 \times n^{loc}_4$ with
$n^{loc}_i = n_i/p_i$ on each processor. Here, for simplicity, we
assume that each $p_i$ divides $n_i$ and that we have $n^{loc}_i \geq
2$ for $i = 1,\ldots,4$.

Let us partition the whole lattice into $n^{loc} = n^{loc}_1 n^{loc}_2
n^{loc}_3 n^{loc}_4$ groups.  Each group corresponds to a fixed
position of the local grid and contains all grid points appearing at
this position within their respective local grid.  Associating a color
with each of the groups, we can interpret this process as a coloring
of the lattice points. For the 2-dimensional case, such a coloring is
depicted in Figure~1 (with $n^{loc}_1 = n^{loc}_2 = 4$), where the 16
colors are denoted by the letters $a$ -- $q$.

We now consider an ordering where all points of color $b$ are ordered after
all those of color $a$ if on the local grids the position belonging
to color $a$ is lexicographically less than that of color $b$. In Figure~1,
this corresponds to the alphabetic ordering of the colors $a$ -- $q$.
Such an ordering is termed {\em locally lexicographic}. 
From now on we will use the prefix `$ll$-'
in expressions like `$ll$-first', `$ll$-ordering' to make clear that they 
refer to the locally lexicographic ordering.

\setlength{\unitlength}{0.7cm}
\begin{figure}
\begin{center}
\begin{picture}(12,10)
\thicklines
\multiput(0,0)(0,1){10}{\multiput(0,0)(1,0){12}{\circle{0.5}}}
\multiput(1.5,-0.5)(4,0){3}{\line(0,1){10}}
\multiput(-0.5,2.5)(0,4){2}{\line(1,0){12}}
\put(2.25,6){\vector(1,0){0.5}}
\put(2,4.75){\vector(0,-1){0.5}}
\put(4.25,5){\vector(1,0){0.5}}
\put(5,5.75){\vector(0,-1){0.5}}
\put(5.75,5){\vector(-1,0){0.5}}
\put(8,4.75){\vector(0,-1){0.5}}
\put(7.25,4){\vector(1,0){0.5}}
\put(4.25,3){\vector(1,0){0.5}}
\put(5,2.25){\vector(0,1){0.5}}
\put(5,3.75){\vector(0,-1){0.5}}
\put(5.75,3){\vector(-1,0){0.5}}
\put(6.25,4){\vector(1,0){0.5}}
\put(7,4.75){\vector(0,-1){0.5}}
\begin{tiny}
\begin{em}
\put(0,9){\makebox(0,0){g}}
\put(1,9){\makebox(0,0){h}}
\put(2,9){\makebox(0,0){e}}
\put(3,9){\makebox(0,0){f}}
\put(4,9){\makebox(0,0){g}}
\put(5,9){\makebox(0,0){h}}
\put(6,9){\makebox(0,0){e}}
\put(7,9){\makebox(0,0){f}}
\put(8,9){\makebox(0,0){g}}
\put(9,9){\makebox(0,0){h}}
\put(10,9){\makebox(0,0){e}}
\put(11,9){\makebox(0,0){f}}
\put(0,8){\makebox(0,0){l}}
\put(1,8){\makebox(0,0){m}}
\put(2,8){\makebox(0,0){i}}
\put(3,8){\makebox(0,0){k}}
\put(4,8){\makebox(0,0){l}}
\put(5,8){\makebox(0,0){m}}
\put(6,8){\makebox(0,0){i}}
\put(7,8){\makebox(0,0){k}}
\put(8,8){\makebox(0,0){l}}
\put(9,8){\makebox(0,0){m}}
\put(10,8){\makebox(0,0){i}}
\put(11,8){\makebox(0,0){k}}
\put(0,7){\makebox(0,0){p}}
\put(1,7){\makebox(0,0){q}}
\put(2,7){\makebox(0,0){n}}
\put(3,7){\makebox(0,0){o}}
\put(4,7){\makebox(0,0){p}}
\put(5,7){\makebox(0,0){q}}
\put(6,7){\makebox(0,0){n}}
\put(7,7){\makebox(0,0){o}}
\put(8,7){\makebox(0,0){p}}
\put(9,7){\makebox(0,0){q}}
\put(10,7){\makebox(0,0){n}}
\put(11,7){\makebox(0,0){o}}
\put(0,6){\makebox(0,0){c}}
\put(1,6){\makebox(0,0){d}}
\put(2,6){\makebox(0,0){a}}
\put(3,6){\makebox(0,0){b}}
\put(4,6){\makebox(0,0){c}}
\put(5,6){\makebox(0,0){d}}
\put(6,6){\makebox(0,0){a}}
\put(7,6){\makebox(0,0){b}}
\put(8,6){\makebox(0,0){c}}
\put(9,6){\makebox(0,0){d}}
\put(10,6){\makebox(0,0){a}}
\put(11,6){\makebox(0,0){b}}
\put(0,5){\makebox(0,0){g}}
\put(1,5){\makebox(0,0){h}}
\put(2,5){\makebox(0,0){e}}
\put(3,5){\makebox(0,0){f}}
\put(4,5){\makebox(0,0){g}}
\put(5,5){\makebox(0,0){h}}
\put(6,5){\makebox(0,0){e}}
\put(7,5){\makebox(0,0){f}}
\put(8,5){\makebox(0,0){g}}
\put(9,5){\makebox(0,0){h}}
\put(10,5){\makebox(0,0){e}}
\put(11,5){\makebox(0,0){f}}
\put(0,4){\makebox(0,0){l}}
\put(1,4){\makebox(0,0){m}}
\put(2,4){\makebox(0,0){i}}
\put(3,4){\makebox(0,0){k}}
\put(4,4){\makebox(0,0){l}}
\put(5,4){\makebox(0,0){m}}
\put(6,4){\makebox(0,0){i}}
\put(7,4){\makebox(0,0){k}}
\put(8,4){\makebox(0,0){l}}
\put(9,4){\makebox(0,0){m}}
\put(10,4){\makebox(0,0){i}}
\put(11,4){\makebox(0,0){k}}
\put(0,3){\makebox(0,0){p}}
\put(1,3){\makebox(0,0){q}}
\put(2,3){\makebox(0,0){n}}
\put(3,3){\makebox(0,0){o}}
\put(4,3){\makebox(0,0){p}}
\put(5,3){\makebox(0,0){q}}
\put(6,3){\makebox(0,0){n}}
\put(7,3){\makebox(0,0){o}}
\put(8,3){\makebox(0,0){p}}
\put(9,3){\makebox(0,0){q}}
\put(10,3){\makebox(0,0){n}}
\put(11,3){\makebox(0,0){o}}
\put(0,2){\makebox(0,0){c}}
\put(1,2){\makebox(0,0){d}}
\put(2,2){\makebox(0,0){a}}
\put(3,2){\makebox(0,0){b}}
\put(4,2){\makebox(0,0){c}}
\put(5,2){\makebox(0,0){d}}
\put(6,2){\makebox(0,0){a}}
\put(7,2){\makebox(0,0){b}}
\put(8,2){\makebox(0,0){c}}
\put(9,2){\makebox(0,0){d}}
\put(10,2){\makebox(0,0){a}}
\put(11,2){\makebox(0,0){b}}
\put(0,1){\makebox(0,0){g}}
\put(1,1){\makebox(0,0){h}}
\put(2,1){\makebox(0,0){e}}
\put(3,1){\makebox(0,0){f}}
\put(4,1){\makebox(0,0){g}}
\put(5,1){\makebox(0,0){h}}
\put(6,1){\makebox(0,0){e}}
\put(7,1){\makebox(0,0){f}}
\put(8,1){\makebox(0,0){g}}
\put(9,1){\makebox(0,0){h}}
\put(10,1){\makebox(0,0){e}}
\put(11,1){\makebox(0,0){f}}
\put(0,0){\makebox(0,0){l}}
\put(1,0){\makebox(0,0){m}}
\put(2,0){\makebox(0,0){i}}
\put(3,0){\makebox(0,0){k}}
\put(4,0){\makebox(0,0){l}}
\put(5,0){\makebox(0,0){m}}
\put(6,0){\makebox(0,0){i}}
\put(7,0){\makebox(0,0){k}}
\put(8,0){\makebox(0,0){l}}
\put(9,0){\makebox(0,0){m}}
\put(10,0){\makebox(0,0){i}}
\put(11,0){\makebox(0,0){k}}
\end{em}
\end{tiny}
\end{picture}
\end{center}
\caption{Locally lexicographic ordering and forward solve in 2 dimensions} 
\end{figure}
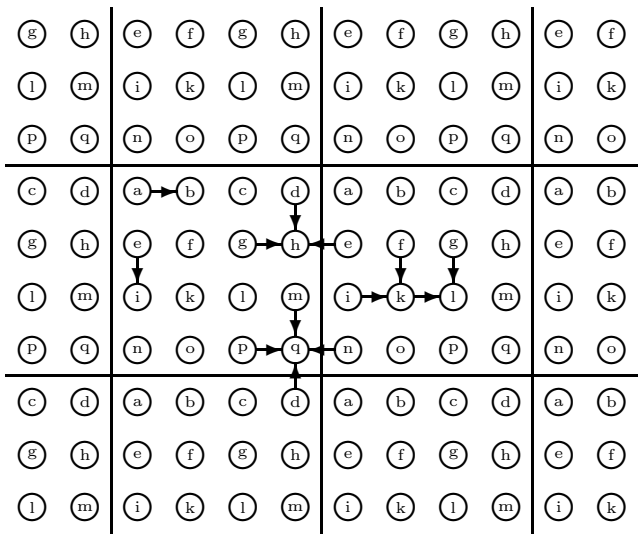

Since we assumed $n^{loc}_i
\geq 2$ for all $i$, all nearest neighbors of a given grid point have 
colors different from that point. This implies that when performing the
forward and backward solves in Algorithm~\ref{SSOR-BiCGstab}, grid points
having the same color can be worked upon in parallel, thus yielding  
an optimal parallelism of 
$p$, the number of processors. A formulation of the 
$ll$-forward solve is given as Algorithm~\ref{ll-forward}. 
Here, we use `$\leq_{ll}$' as a symbol for `$ll$-less than'.

\begin{algor}[htb]
\begin{center}
\begin{minipage}{\textwidth}
\begin{tabbing}
\hspace*{2ex} \= \hspace{2ex} \=  \hspace{2ex} \= \hspace{2ex} \kill
for all colors in lexicographic order \\
\> for all processors \\
\>  \> $x := $ grid point of that color on that processor \\
\>  \> \{ update $y_x$ \} \\
\>  \>  $\displaystyle y_x = p_x + \kappa \left( 
         \sum_{\mu, \, x-\mu \, \leq_{ll} \, x} m^+_{x,x-\mu}y_{x - \mu}
         + \sum_{\mu, \, x+\mu \, \leq_{ll} \, x} m^-_{x,x+\mu}y_{x+\mu}
              \right)$
\end{tabbing}
\end{minipage}
\end{center}
\caption{$ll$-forward solve\label{ll-forward}}
\end{algor}

For grid points lying in the `interior' of
each local grid, we have $x-\mu \leq_{ll} x \leq_{ll} x+\mu$ 
for $\mu = 1,\ldots,4$.
The update thus becomes
\[
y_x = p_x + \kappa \left(
         \sum_{\mu= 1}^{4} m^+_{x,x-\mu}y_{x - \mu} \right),   
\]
whereas on the `local boundaries' we will have between 0 (for the
$ll$-first point) and 8 (for the $ll$-last point) summands to add to
$p_x$.  The second sum in the update formula then contains those grid
points which belong to neighboring processors. The arrows in Figure~1
illustrate this situation for the 2-dimensional analogue, where 2 grid
points contribute to the update at locally interior points (colors
$f,g,k,l$) and 0 to 4 on the local boundaries. (An arrow from point
$x\pm \mu$ to $x$ means that $x\pm \mu$ appears in the sum for
updating point $x$.)

With the lexicographic ordering on the {\em whole} lattice, a forward
solve followed by a backward solve has the effect that the information
at any grid point will have spread out to all other grid points. This
yields a heuristic justification of why this ordering is superior (as
far as convergence properties are concerned) to the odd-even ordering,
where the information is passed to the (second-) nearest neighbors
only. The $ll$-ordering can be regarded as a compromise between these
two extreme cases, spreading information to all points within a local
grid. The $ll$-forward and $ll$-backward solves parallelize
efficiently and better than with the lexicographic ordering on the
whole grid.  The parallelism achieved is $p$ and thus less than with
the odd-even ordering, but it is optimal since we have $p$ processors.
If we change the number of processors, the $ll$-ordering, and
consequently the properties of the corresponding SSOR preconditioner
will change, too. Heuristically, we expect the convergence properties
to degrade as the size of the local grid becomes smaller but to always
remain better than with the odd-even preconditioner (as long as
$n_i^{loc} \geq 2$ for $i=1,\ldots,4$).

\section{Parallel Implementation\label{IMPLEMENTATION}}

In this section we give a detailed discussion on the realization of
the $ll$-forward solve on a parallel computer. In order to render all
necessary communication transparent we assume a model parallel
computer where communication is done via explicit message passing
between processors.

As in the previous section, we assume that the processors are
connected as a 4-dimensional (or lower dimensional) grid and that the
space-time lattice is mapped onto that grid in the obvious manner,
yielding a local subgrid on each processor.  The local coordinates of
a grid point $x$ within a subgrid are denoted by the 4-tuple $x =
(i_1,i_2,i_3,i_4)$. For processor $\pi$ of the processor grid we
denote its neighbors in dimension $\mu$ by $\pi\pm \mu$.
Incorporating all necessary data transmission into
Algorithm~\ref{ll-forward} via message passing statements---sending
messages as soon as possible while receiving them as late as
possible---we end up with Algorithm~\ref{ll-forward-messages} which
formulates the $ll$-forward solve for processor $\pi$.

\begin{algor}[htb]
\begin{center}
\begin{minipage}{\textwidth}
\begin{tabbing}
\hspace*{2ex} \= \hspace{2ex} \=  \hspace{2ex} \= \hspace{2ex} \= \hspace{2ex}
   \= \hspace{2ex} \= \kill
for $i_4 = 1,\ldots,n_4^{loc}$  \\
\> for $i_3 = 1,\ldots,n_3^{loc}$  \\
\>  \> for $i_2 = 1,\ldots,n_2^{loc}$  \\
\>  \>  \> for $i_1 = 1,\ldots,n_1^{loc}$  \\
\>  \> \> \> $x := (i_1,i_2,i_3,i_4)$ \\
\>  \> \>  \> \{ update $y_x$ \} \\
\>  \> \>  \> $y_x = p_x $ \\
\>  \> \>  \> for $\mu = 1,\ldots,4$ \\
\>  \> \> \>  \> if $i_{\mu} > 1$ then $y_x = y_x + 
                        \kappa m^{+}_{x,x-\mu}y_{x-\mu}$  \\
\>  \> \>  \> for $\mu = 1,\ldots,4$ \\
\>  \> \>  \> \> if $i_{\mu} = 1$ then \\
\>  \> \>  \> \>  \> send $y_x$ to processor $\pi-\mu$ \\
\>  \> \>  \> \> if $i_{\mu} = n_{\mu}^{loc}$ then \\
\>  \> \>  \> \> \>  receive $y_{x+\mu}$ from processor $\pi+\mu$ \\  
\>  \> \>  \> \> \> $y_x = y_x + \kappa m^{-}_{x,x+\mu}y_{x+\mu}$      
\end{tabbing}
\end{minipage}
\end{center}
\caption{$ll$-forward solve on  processor $\pi$\label{ll-forward-messages}}
\end{algor}

Note that if processor $\pi$ issues a `send' to transfer $y_x$ to
processor $\pi-\mu$ because of $i_{\mu} = 1$, the corresponding
`receive' (for $i_{\mu} = n_{\mu}^{loc}$) in processor $\pi-\mu$ reads
this data as $y_{x'}$ with $x'=x+ \mu$. If $\mu =1$, exactly
$n_1^{loc}-2$ updates are done between a matching send/receive pair.
If $\mu = 2$ this number increases to $n^{loc}_1(n_2^{loc}-2)$, for
$\mu = 3$ it becomes $n^{loc}_1n^{loc}_2(n^{loc}_3 - 2)$ and for $\mu
= 4$ it is $n^{loc}_1n^{loc}_2n^{loc}_3(n^{loc}_4 - 2)$. Therefore, if
messages can be handled in parallel to the computation, the message
passing can be hidden behind the computation so that the communication
overhead will be small. Also note that if $p_{\mu} = 1$ the
send/receive pairs for direction $\mu$ have to be dropped in
Algorithm~\ref{ll-forward-messages}.

Algorithm~\ref{ll-forward-messages} has to communicate the local
boundary values all one by one. This is inherent in the $ll$-ordering
and can become a disadvantage on machines which have a large start-up
time for communication, irrespective of the message length. In these
cases, it is preferable to issue only a small number of (large)
messages. Then the following  
outer product' of the $ll$-ordering with
an odd-even ordering can be used: Divide each local grid into two
halves $G_1, G_2$ by bisecting perpendicularly to direction 4 (say),
where $n^{loc}_4$ is supposed to be a multiple of 2.  Identify odd and
even processors on the processor grid, neglecting the 4-th coordinate,
i.e., processor $\pi = (j_1,j_2,j_3,j_4)$ is odd if and only if $j_1 +
j_2 + j_3$ is odd.  On each even processor, we first take a locally
lexicographic ordering on $G_1$ and then on $G_2$, on odd processors
we reverse the role of $G_1$ and $G_2$. The forward solve with respect
to this ordering consists of two half steps.  First, the odd
processors update $G_1$ while the even processors work on $G_2$. No
communication is necessary there. Then the updated values on the local
boundaries have to be communicated (requiring a total of 7 large
messages, one for direction 4 and two for each other direction). Now
the second half step can be performed, where odd processors update
$G_2$ and even processors update $G_1$. An illustration of this $oe
\times ll$--ordering for the 2-dimensional case is given in
Figure~\ref{oexll}.

\setlength{\unitlength}{0.7cm}
\begin{figure}
\begin{center}
\begin{picture}(12,10)
\thicklines
\multiput(0,0)(0,1){10}{\multiput(0,0)(1,0){12}{\circle{0.5}}}
\multiput(1.5,-0.5)(4,0){3}{\line(0,1){10}}
\multiput(-0.5,2.5)(0,4){2}{\line(1,0){12}}
\multiput(-0.5,0.5)(0,4){3}{\multiput(0,0)(0.5,0){25}{\line(1,0){0.25}}}
\put(2.25,6){\vector(1,0){0.5}}
\put(2.25,5){\vector(1,0){0.5}}
\put(3,5.75){\vector(0,-1){0.5}}
\put(4.25,5){\vector(1,0){0.5}}
\put(5,5.75){\vector(0,-1){0.5}}
\put(8,6.75){\vector(0,-1){0.5}}
\put(7.25,6){\vector(1,0){0.5}}
\put(7.25,4){\vector(1,0){0.5}}
\put(4.25,3){\vector(1,0){0.5}}
\put(5,2.25){\vector(0,1){0.5}}
\put(5,3.75){\vector(0,-1){0.5}}
\put(5.75,3){\vector(-1,0){0.5}}
\put(2.25,3){\vector(1,0){0.5}}
\put(3,3.75){\vector(0,-1){0.5}}
\put(3,2.25){\vector(0,1){0.5}}
\begin{tiny}
\begin{em}
\put(0,9){\makebox(0,0){p}}
\put(1,9){\makebox(0,0){q}}
\put(2,9){\makebox(0,0){e}}
\put(3,9){\makebox(0,0){f}}
\put(4,9){\makebox(0,0){g}}
\put(5,9){\makebox(0,0){h}}
\put(6,9){\makebox(0,0){n}}
\put(7,9){\makebox(0,0){o}}
\put(8,9){\makebox(0,0){p}}
\put(9,9){\makebox(0,0){q}}
\put(10,9){\makebox(0,0){e}}
\put(11,9){\makebox(0,0){f}}
\put(0,8){\makebox(0,0){c}}
\put(1,8){\makebox(0,0){d}}
\put(2,8){\makebox(0,0){i}}
\put(3,8){\makebox(0,0){k}}
\put(4,8){\makebox(0,0){l}}
\put(5,8){\makebox(0,0){m}}
\put(6,8){\makebox(0,0){a}}
\put(7,8){\makebox(0,0){b}}
\put(8,8){\makebox(0,0){c}}
\put(9,8){\makebox(0,0){d}}
\put(10,8){\makebox(0,0){i}}
\put(11,8){\makebox(0,0){k}}
\put(0,7){\makebox(0,0){g}}
\put(1,7){\makebox(0,0){h}}
\put(2,7){\makebox(0,0){n}}
\put(3,7){\makebox(0,0){o}}
\put(4,7){\makebox(0,0){p}}
\put(5,7){\makebox(0,0){q}}
\put(6,7){\makebox(0,0){e}}
\put(7,7){\makebox(0,0){f}}
\put(8,7){\makebox(0,0){g}}
\put(9,7){\makebox(0,0){h}}
\put(10,7){\makebox(0,0){n}}
\put(11,7){\makebox(0,0){o}}
\put(0,6){\makebox(0,0){l}}
\put(1,6){\makebox(0,0){m}}
\put(2,6){\makebox(0,0){a}}
\put(3,6){\makebox(0,0){b}}
\put(4,6){\makebox(0,0){c}}
\put(5,6){\makebox(0,0){d}}
\put(6,6){\makebox(0,0){i}}
\put(7,6){\makebox(0,0){k}}
\put(8,6){\makebox(0,0){l}}
\put(9,6){\makebox(0,0){m}}
\put(10,6){\makebox(0,0){a}}
\put(11,6){\makebox(0,0){b}}
\put(0,5){\makebox(0,0){p}}
\put(1,5){\makebox(0,0){q}}
\put(2,5){\makebox(0,0){e}}
\put(3,5){\makebox(0,0){f}}
\put(4,5){\makebox(0,0){g}}
\put(5,5){\makebox(0,0){h}}
\put(6,5){\makebox(0,0){n}}
\put(7,5){\makebox(0,0){o}}
\put(8,5){\makebox(0,0){p}}
\put(9,5){\makebox(0,0){q}}
\put(10,5){\makebox(0,0){e}}
\put(11,5){\makebox(0,0){f}}
\put(0,4){\makebox(0,0){c}}
\put(1,4){\makebox(0,0){d}}
\put(2,4){\makebox(0,0){i}}
\put(3,4){\makebox(0,0){k}}
\put(4,4){\makebox(0,0){l}}
\put(5,4){\makebox(0,0){m}}
\put(6,4){\makebox(0,0){a}}
\put(7,4){\makebox(0,0){b}}
\put(8,4){\makebox(0,0){c}}
\put(9,4){\makebox(0,0){d}}
\put(10,4){\makebox(0,0){i}}
\put(11,4){\makebox(0,0){k}}
\put(0,3){\makebox(0,0){g}}
\put(1,3){\makebox(0,0){h}}
\put(2,3){\makebox(0,0){n}}
\put(3,3){\makebox(0,0){o}}
\put(4,3){\makebox(0,0){p}}
\put(5,3){\makebox(0,0){q}}
\put(6,3){\makebox(0,0){e}}
\put(7,3){\makebox(0,0){f}}
\put(8,3){\makebox(0,0){g}}
\put(9,3){\makebox(0,0){h}}
\put(10,3){\makebox(0,0){n}}
\put(11,3){\makebox(0,0){o}}
\put(0,2){\makebox(0,0){l}}
\put(1,2){\makebox(0,0){m}}
\put(2,2){\makebox(0,0){a}}
\put(3,2){\makebox(0,0){b}}
\put(4,2){\makebox(0,0){c}}
\put(5,2){\makebox(0,0){d}}
\put(6,2){\makebox(0,0){i}}
\put(7,2){\makebox(0,0){k}}
\put(8,2){\makebox(0,0){l}}
\put(9,2){\makebox(0,0){m}}
\put(10,2){\makebox(0,0){a}}
\put(11,2){\makebox(0,0){b}}
\put(0,1){\makebox(0,0){p}}
\put(1,1){\makebox(0,0){q}}
\put(2,1){\makebox(0,0){e}}
\put(3,1){\makebox(0,0){f}}
\put(4,1){\makebox(0,0){g}}
\put(5,1){\makebox(0,0){h}}
\put(6,1){\makebox(0,0){n}}
\put(7,1){\makebox(0,0){o}}
\put(8,1){\makebox(0,0){p}}
\put(9,1){\makebox(0,0){q}}
\put(10,1){\makebox(0,0){e}}
\put(11,1){\makebox(0,0){f}}
\put(0,0){\makebox(0,0){c}}
\put(1,0){\makebox(0,0){d}}
\put(2,0){\makebox(0,0){i}}
\put(3,0){\makebox(0,0){k}}
\put(4,0){\makebox(0,0){l}}
\put(5,0){\makebox(0,0){m}}
\put(6,0){\makebox(0,0){a}}
\put(7,0){\makebox(0,0){b}}
\put(8,0){\makebox(0,0){c}}
\put(9,0){\makebox(0,0){d}}
\put(10,0){\makebox(0,0){i}}
\put(11,0){\makebox(0,0){k}}
\end{em}
\end{tiny}
\end{picture}
\end{center}
\caption{$ll \times oe$ ordering and forward solve, 2 dimensions\label{oexll}} 
\end{figure}
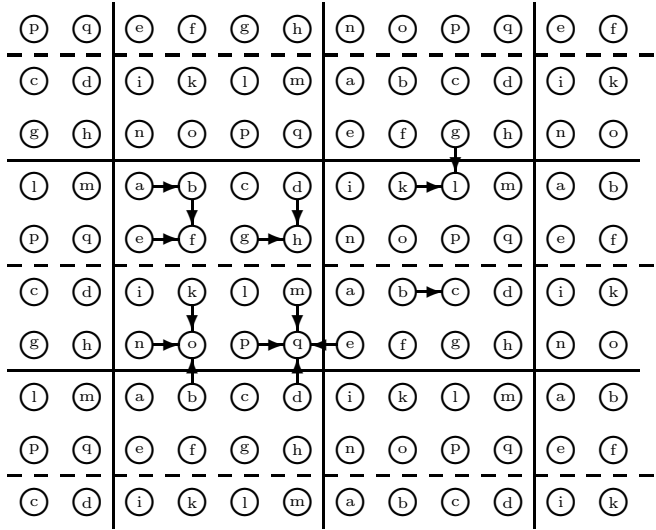

Implementing the $ll$-SSOR preconditioning requires to 
distinguish between grid sites along
a given direction. Assume in a forward solve that a grid point is
located at the `left' boundary of the local lattice. This site
is {\em not} updated with data from the left neighbor, and
this fact has to be inquired by a conditional statement or a case
statement.  The second case is that a grid site is located at a
`right' boundary where the right neighbor has to be communicated
from the neighbor sublattice (in this situation the neighbor
processor).  The third possibility is that the site is not a boundary
grid point. This leads to 81 different cases a site can be associated
with in four dimensions.

\section{Results\label{RESULTS}}

Our numerical tests of the locally lexicographic SSOR preconditioner
were performed on APE100/Quadrics machines, a SIMD parallel
architecture optimized for fast floating point arithmetic on block
data-structures like $3\times 3$ SU(3) matrices or any other data
structure that can be blocked similarly. Arithmetic on the
Quadrics is compliant with the IEEE 754 single precision standard.
We had access to a 32-node
Quadrics Q4 at Wuppertal University and a 512-node Quadrics QH4 at the
computer center of ENEA/Casaccia Rome.

In order to speed up the performance on the Quadrics it is favorable
(as it is the case on most high speed parallel computers) to partition
the code into code blocks \ie\ sections that can be computed entirely
from the registers without recourse to the local or remote memory,
thus using the pipelining features of the processors efficiently.
This requires to modify Algorithm~\ref{ll-forward-messages} by moving
all if-statements out of the four-fold nested loop over
$i_1,\ldots,i_4$ resulting in a sequence of 81 nested loops, one for
each possible combination of the three cases $i_{\mu} = 1, 1 < i_{\mu}
< n_{\mu}^{loc}, i_{\mu} = n_{\mu}^{loc}$ for $\mu = 1,\ldots,4$.

\begin{figure}[b]
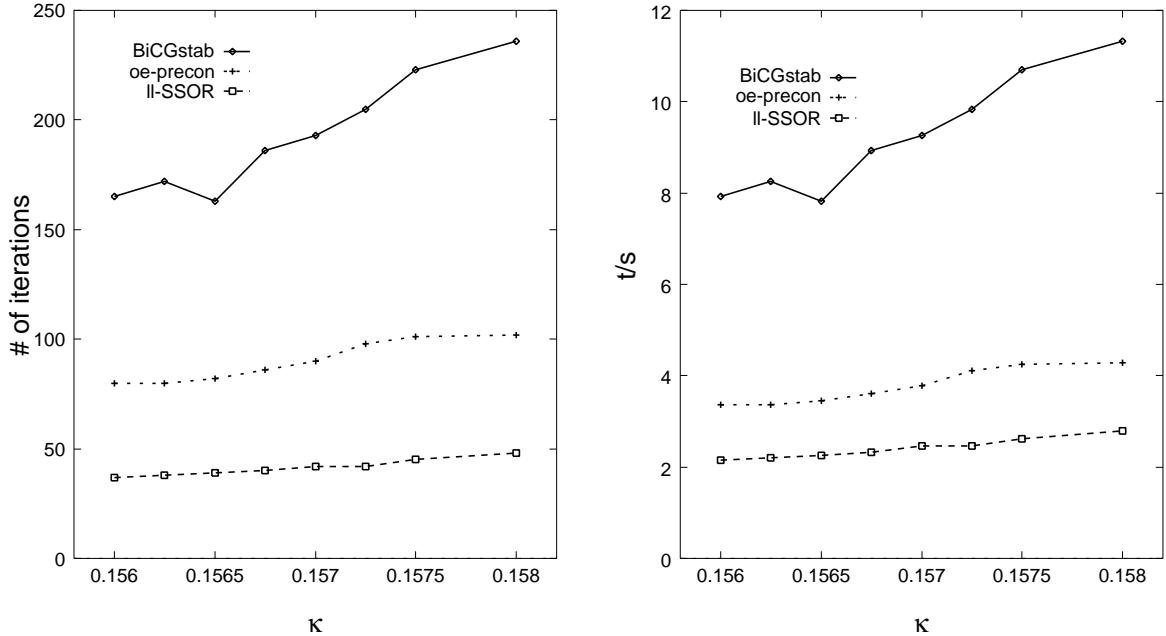

\centerline{\epsfxsize=.48\textwidth\epsfbox{sea-val.eps}\hfill
            \epsfxsize=.48\textwidth\epsfbox{sea-val-time.eps}}
\caption{Iteration numbers of pure BiCGstab $\bullet$, the  $oe$- $+$ and the 
  $ll$-preconditioned version $\Box$ for a series of valence masses on a
  $8^3\times 16$ lattice. The second plot shows the cpu-time needed on
  a 32-node Quadrics Q4 at granularity $g=256$.\label{VALENCE}}
\end{figure}

\paragraph{Evidence of improvement.}
In Fig.~\ref{VALENCE} first evidence of improvement by $ll$-SSOR
preconditioning of BiCGstab is presented.
Shown are three different iteration numbers of a valence mass
``trajectory'' for a given thermalized full-QCD configuration at
$\beta=5.6$ and $\kappa_{\mbox{\tiny sea}}=0.156$ corresponding
 to an intermediate pion
mass.  The standard BiCGstab solution of \eqnref{Wilson_eq} is compared
with the odd-even preconditioned method and the new locally
lexicographic ordering scheme.  The measurements are taken on a
lattice of size $8^3\times 16$, hence the computational problem being
of granularity $g=N/p=256$ on the Quadrics Q4 with 32 nodes.  In this
exploratory implementation, as discussed in the previous sections,
the sublattice administrated by a given
processor is used as the local lattice in the $ll$-SSOR scheme.  At
this place, the convergence stopping criteria, \beq \mbox{stop
iteration if}\qquad \frac{||M x -\phi||_{2}}{||\phi||_{2}} < \epsilon, \eeq
are applied on the three different residuals used in the three variant
algorithms, respectively, with equal numerical value
$\epsilon=10^{-8}$. We verified that the `true' residuals $Mx - \phi$,
calculated explicitly from
the solution $x$, eventually were smaller for the preconditioned
variants of the BiCGstab solver.  The gain in iterations is more than
a factor of 2 for $\kappa_{\mbox{\tiny valence}} = 
\kappa_{\mbox{\tiny sea}}$ as is the case within the entire valence
$\kappa$ range considered.  Compared to the unpreconditioned BiCGstab
the gain is even a factor of 4.

The improvement in terms of cpu-time, on the Q4, is smaller than the
gain in iteration numbers since the $ll$-SSOR preconditioning
leads to certain breaks in code blocks due to the 81 different cases
mentioned above. Thus, pipelined arithmetic is not exploited as well
as in the other cases. For the given choice of parameters we find an
improvement of about a factor of about 1.5 compared to standard
$oe$-preconditioning throughout the valence mass range considered.

\begin{figure}[htb]
\centerline{\epsfxsize=.5\textwidth\epsfbox{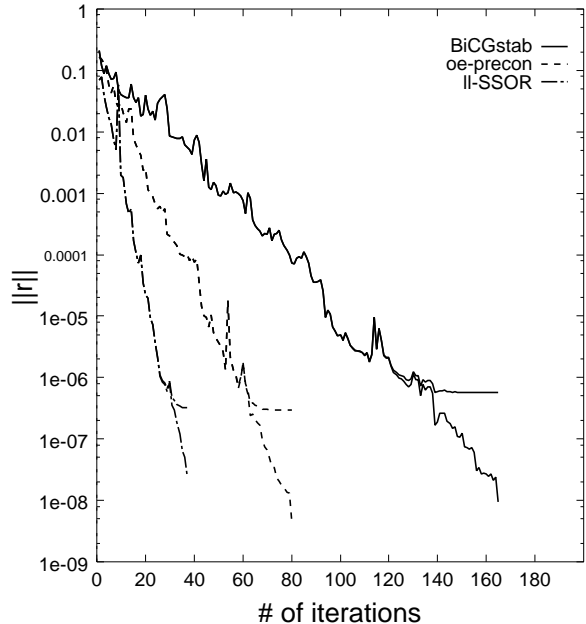}}
\caption{Convergence behavior of BiCGstab and its $oe$- and 
  $ll$-preconditioned variants on an $8^3\times 16$
  lattice.\label{CONVERGENCE}}
\end{figure}

The convergence behavior of the three methods is demonstrated in
Fig.~\ref{CONVERGENCE} on a typical configuration at the same
parameter values for $\kappa_{\mbox{\tiny sea}}=\kappa_{\mbox{\tiny
    val}}=0.156$.
In addition to the acceleration, the convergence behavior turns out to be
much smoother for the $ll$-preconditioned BiCGstab than in the
unpreconditioned case. We have plotted both the accumulated and the
`true' residual. The latter approaches a horizontal line for
$\|r\|_2 \approx 10^{-6}$ which indicates
the limits of single precision accuracy. For
$ll$-preconditioning, however, the final residual that can be achieved
appears to be somewhat more accurate than for pure BiCGstab.

\paragraph{Local lattice size dependency.}
It is important to assess the dependency of the improvement due to
$ll$-SSOR preconditioning on the size of the local lattice.  As our Q4
is equipped with a fixed number of 32 processors, we emulated larger numbers of
local lattices by sub-blocking the system on each processor.  The smallest
local lattice size we can investigate is a $2^4$ lattice, apart from
the unpreconditioned case (which we treat as having local lattice size 1)
and the $oe$-realization of the solution of
\eqnref{Wilson_eq} (local lattice size 2).

\begin{figure}[htb]
\centerline{\epsfxsize=.48\textwidth\epsfbox{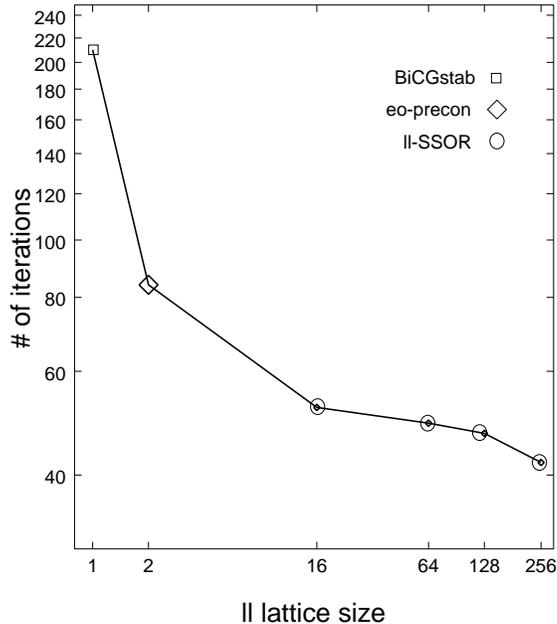}}
\caption{Dependence of iteration numbers 
  on the $ll$-lattice size, the lattice volume is $8^3\times
  16$.\label{SATURATION}}
\end{figure}

The dependency of the number of iterations on the local lattice size
is depicted in Fig.~\ref{SATURATION} on an equally sized lattice as
used before and at equal physical parameters but with
$\kappa_{\mbox{\tiny sea}}=0.1575$.
As expected, smaller local lattices lead to a less efficient
preconditioning.  However, the number of iterations at first increases only
slightly if we move from a local lattice size of $256$ to
$16$. This behavior might of course be dependent on
$\kappa_{\mbox{\tiny sea}}$, hence the crossover to the $oe$-result
(the diamond in Fig.~\ref{SATURATION}) might occur at a larger local
lattice size for $\kappa_{\mbox{\tiny sea}}$
closer to $\kappa_{c}$.

\begin{figure}[htb]
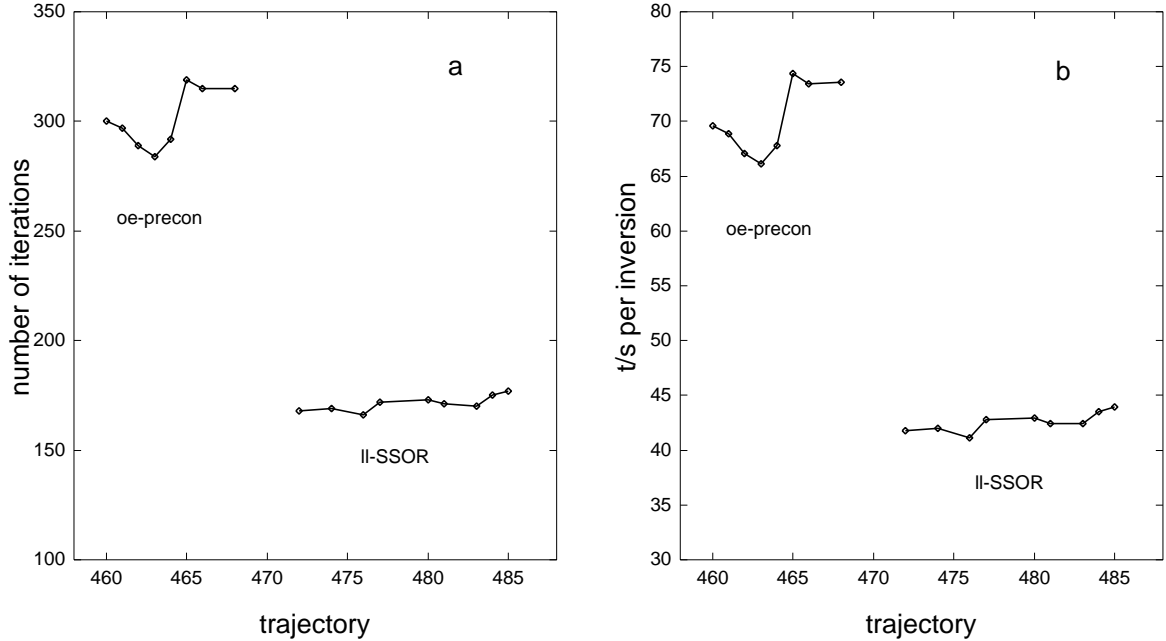

\centerline{\epsfxsize=.48\textwidth\epsfbox{full.eps}\hfill
            \epsfxsize=.48\textwidth\epsfbox{full-time.eps}}
\caption{Section of the time series of iteration numbers together
  with cpu-times required on a QH4 from a full QCD HMC simulation on a
  $24^3\times 40$ lattice\label{FULL}.  The graphics shows the
  location where we switched from the $oe$-representation of the
  fermionic matrix $M$ to the $ll$-scheme.}
\end{figure}

\paragraph{Speeding up QCD computations.}
Next we discuss improvements due to the $ll$-SSOR preconditioning
implemented in an actual
Hybrid Monte Carlo simulation of full QCD with Wilson fermions. We
depict in Fig.~\ref{FULL}a a time series of iteration numbers that are
each averages over a molecular dynamics trajectory of a length of 125
steps. The measurements are taken on a system of size $24^3\times 40$
at $\beta=5.6$ and $\kappa_{\mbox{\tiny sea}}=0.1575$.
The first part of the curve has been generated using standard
$oe$-preconditioning for the BiCGstab inverter.  The sudden step in
iterations is due to the fact that we switched the preconditioner to
$ll$-SSOR at this location. The iteration numbers show a quite stable
value in both branches of the curve and hence nicely demonstrate the
stability of the preconditioning on different background field
configurations within the canonical ensemble. In the actual large
scale simulation, the $ll$-SSOR preconditioning scheme indeed gives an
improvement of nearly a factor of 2 as to the iteration numbers for a
quite light pion mass \cite{SESAM95}.

We remark that we have verified in our simulation that the educated
guess procedure as advocated by Brower et al.\ \cite{Br95} leads to an
improvement that is {\em additive} to the gain as achieved by the use
of $ll$-SSOR preconditioning.

The time behavior of the system on the 512-node APE100/Quadrics QH4
is depicted in Fig.~\ref{FULL}b. We remark that on this machine
topology with $8\times 8\times 8$ nodes on a 3-dimensional grid, we
arrive at a local grid size of $3\times 3 \times 24 \times 5$ for the
$24^3\times 40$ lattice. On a strict SIMD architecture without local
addressing, as is the case on the Quadrics, $oe$-schemes are quite
tricky to implement and lead to a substantial loss in performance in
the range of 30 \%.  Therefore $ll$-SSOR can help a lot to overcome the SIMD
odd-even limitations.  The total improvement as to cpu-time is about a
factor of 1.7.

Details on our implementation and the experience with $ll$-SSOR
preconditioning in the context of large scale HMC simulations in full
QCD will be published elsewhere \cite{SESAM96}.

\section{Conclusion and Outlook}

We have presented a new local grid point ordering scheme that allows
to carry through efficient preconditioning of Krylov subspace solvers
like BiCGstab in the context of computations of Greens functions for
lattice quantum chromodynamics with Wilson fermions on parallel
computers.

Using the Eisenstat-trick we can avoid the explicit multiplication by
the fermionic matrix $M$ and we remain with  a
forward and a backward solve to invert the upper and lower triangular
parts of $M$.

In actual Hybrid Monte Carlo simulations with Wilson fermions at small
pion masses and on large lattices, we verify an improvement of a
factor of 2 in BiCGstab concerning iteration numbers as compared to the
state-of-the-art odd-even preconditioned scheme. In this manner, our
code is thus speeded up on a 512-node APE100/Quadrics by a factor of 
about 1.7 at $\beta=5.6$ and $\kappa_{\mbox{\tiny sea}}=0.1575$. 

We are confident that the $ll$-SSOR scheme will help to drive future
simulations with dynamical Wilson fermions deeper into the chiral
region dominated by a small pion mass.

\section*{Acknowledgments}

Part of the numerical test in this paper have been carried out on the
Quadrics QH2 at ENEA/Casaccia/Roma.  Th.\ L.\ and G.\ R.\ thank
Profs.\ N.\ Cabibbo and A.\ Mathis for their kind hospitality and
Dotts.\ R.\ Cannata and R.\ Guadagni for very valuable support.
We are grateful to our friends from the SESAM group for important
discussions and their patience while we were blocking the Quadrics Q4
at Wuppertal University and HLRZ/KFA-J\"ulich.
We acknowledge helpful discussions with P. Fiebach from the department
of Mathematics, University of Wuppertal and
G. Hockney from Fermi-Lab, Batavia, IL.

The work of Thomas Lippert is supported by the Deutsche
Forschungsgemeinschaft DFG under grant No.\ Schi 257/5-1.

\end{document}